\documentclass[prl,aps,showpacs,superscriptaddress,prl,amsmath,twocolumn,floatfix]{revtex4}
\usepackage{graphicx}
\newcommand{\eq}[1]{Eq.~(\ref{#1})}
\begin{document}
\title{Van der Waals Density Functional for General Geometries}
\author{M. Dion}		\affiliation{Center for Materials Theory, 
Department of Physics and Astronomy, 
Rutgers University, Piscataway, NJ 08854-8019}
\author{H. Rydberg}		\affiliation{Department of Applied Physics, 
Chalmers University of Technology and 
G{\"o}teborg University,  
SE-412\;96 G\"{o}teborg, Sweden}
\author{E. Schr\"oder}		\affiliation{Department of Applied Physics, 
Chalmers University of Technology and 
G{\"o}teborg University,  
SE-412\;96 G\"{o}teborg, Sweden}
\author{D. C. Langreth}		\affiliation{Center for Materials Theory, 
Department of Physics and Astronomy, 
Rutgers University, Piscataway, NJ 08854-8019}
\author{B. I. Lundqvist}	\affiliation{Department of Applied Physics, 
Chalmers University of Technology and 
G{\"o}teborg University,  
SE-412\;96 G\"{o}teborg, Sweden}

\date{30 January 2004}

\begin{abstract}
A scheme within density functional theory is proposed that provides a practical 
way to generalize to   
unrestricted
geometries the method applied 
with some success to layered geometries 
[H. Rydberg, {\it et al.}, Phys.\ Rev.\ Lett.~\textbf{91}, 126402 (2003)].
It includes van der Waals forces in a seamless fashion. 
By 
expansion 
to second order in 
a carefully chosen quantity contained in 
the long range part of the correlation functional, 
the nonlocal correlations are expressed in terms of a density-density interaction 
formula. It contains a relatively simple parametrized kernel, with 
parameters determined by the local density and its gradient. The proposed functional 
is applied to rare gas and benzene dimers, where it is shown to give a realistic 
description. 
\end{abstract}


\maketitle

Density functional theory (DFT) 
 for  molecules and materials is widely applied with
approximate \emph{local} and
\emph{semilocal} density functionals for the interaction effects. 
For largely homogeneous systems, 
for example, simple metals and semiconductors, 
the local-density approximation (LDA) for these effects is appropriate. 
For inhomogeneous systems,
for example, transition metals, ionic crystals, compound metals,
surfaces, interfaces,
and some chemical systems,
semilocal-density approximations, 
like members of the 
family of
generalized gradient 
approximations (GGA),  
work well.
Today DFT describes cohesion, bonds, structures, 
and other properties very well for dense molecules and materials,
as shown by recent studies for both single molecules \cite{GGAmolecule}
and dense solid-state \cite{GGAsolid} systems.
However, 
sparse systems, including soft matter, van der Waals complexes, and 
biomolecules, are at least as abundant. They have interparticle separations, 
for which 
\emph{nonlocal}, long-ranged interactions, 
such as van der Waals (vdW) forces, 
are 
influential.

The aim of this letter is 
to develop and apply a van der Waals density functional
(vdW-DF) for general geometries to supplement the planar vdW-DF 
that we recently
applied with some success \cite{rydbergPRL} to several layered materials.  
The simplest
form for the 
nonlocal correlation-energy 
part to such a functional is
\begin{equation}
E_\text{c}^\text{nl} = 
\frac{1}{2} \int\!d^3r\,d^3r'\, n(\vec r) 
\phi(\vec r,\vec r\,') n(\vec r\,'),
\label{twopoint}
\end{equation}
where 
$\phi(\vec r,\vec r\,')$ is some given, general function depending on
$\vec r -\vec r\,'$ and the densities $n$ in the vicinity of $\vec r$
and $\vec r\,'$.
 It is approximately derived here and applied to some key cases, with results 
that give promise for broader applications.

We start with the same approximation scheme used for layered systems 
\cite{rydbergPRL,langrethIJQC}, and divide the correlation energy into two pieces,
\begin{equation}
E_{\rm c}[n]=E_{\rm c}^0[n]+E_{\rm c}^{\rm nl}[n],
\label{ec0}
\end{equation}
which are treated in different approximations.  In particular,
as discussed in Ref.~\onlinecite{langrethIJQC}, we treat the
second term in the full potential approximation (FPA), which
is exact at long distances
between separated fragments, 
 and therefore adopt  Eq.~(25) of
Ref.~\onlinecite{langrethIJQC} ,
\begin{equation}
E_\text{c}^\text{nl} =  \int_{0}^{\infty}\frac{du}{2\pi}
{\rm tr}\left[\ln(1-V\tilde\chi) - \ln\epsilon\right],
\label{ediff2}
\end{equation}
where $\tilde\chi$ is the density response to a fully self-consistent potential 
with long-range, 
interfragment spectator \cite{spectator}
contributions omitted \cite{langrethIJQC}.
$V$ is the interelectronic Coulomb interaction, 
$\epsilon$ 
an appropriately approximated dielectric function, and $u$ 
the imaginary frequency. While \eq{ediff2} is taken to be the definition of 
$E_\text{c}^\text{nl}$, we will show that \eq{twopoint} can be obtained with 
suitable approximations.  The first term $E_\text{c}^0$,
defined by Eqs.~(\ref{ec0}) and (\ref{ediff2}),
 is also nonlocal; however, with the long-range vdW terms treated separately, 
it seems a reasonable approximation to treat $E_\text{c}^0$ in the LDA, and 
this is what we do. There is no double counting, because for a uniform system
$1-V\tilde\chi=\epsilon$, which implies that the LDA for $E_\text{c}^\text{nl}$ 
vanishes. This is also the key to a seamless theory.
 
For 
layered systems the scheme 
was made tractable by  
use of the lateral average
of
the densities 
to calculate the interplanar contribution from \eq{ediff2}%
. For general geometries, we make the scheme tractable by expanding 
\eq{ediff2} to second order in $S\equiv 1-\epsilon^{-1}$%
, obtaining 
\begin{equation}
E_\text{c}^\text{nl} \approx \int_0^\infty \frac{du}{4\pi}
\text{tr}{\left[S^2
-\left(\frac{\nabla S \cdot\nabla V}{4\pi e^2}\right)^2\right]}.
\label{sexpand}
\end{equation}
 This vanishes in the uniform limit as it must.  

To assess the nature of this approximation,
we consider its prediction for the long-range interaction between
two fragments.  As in 
 Ref.~\onlinecite{ylvaPRL}, we obtain the standard formula for the coefficient of the 
$R^{-6}$ term in terms of the frequency dependent polarizability tensor 
$\alpha_{ij}^p(\omega)$  for each fragment $p=1,2$.  We find that  
$
\alpha_{ij}^p = 
\delta_{ij}\! \int\! d^3x\!\int\! d^3y\,S^p(\vec x, \vec y)/4\pi.
$
We note that
$S(\vec x, \vec  y)/4\pi$ is the polarization $P$ at point $\vec x$ induced
by the $D$-field at point $\vec y$, 
\textit{i.e.}, $\vec D(\vec y)$ (in this case the field from fluctuations on the
other fragment).  The integration over $x$ then gives 
the dipole moment, and the integration over $y$ implies that this 
 $D$-field is assumed to be spatially constant.
   This means that, although a local-field
correction is included, it is not calculated self-consistently.
It is the same approximation for the electrodynamics that was 
used to calculate the  $C_6$ coefficient successfully over a large 
set of atomic and molecular pairs \cite{ylvaPRL,ylvacomplexes}, an 
approximation also independently developed by another group 
\cite{dobson:ylvafollow}. Fully consistent electrodynamics gave a 
perceptible, but not dramatic,
 statistical improvement \cite{unified} for those atomic 
and molecular dimers tested.  However, one 
should use an improved approximation in cases where the fragment
polarizabilities have significant anisotropy and also for the asymptotic 
interaction between two parallel surfaces, which is not given exactly by this 
approximation.  In future versions of the functional, it will 
probably be possible to eliminate this $S$-expansion, keeping only the expansion 
in terms of the {\it difference} between $\epsilon$ and $1-V\tilde\chi$.  
The retention of the anisotropic components of 
{\boldmath{$\epsilon$}} should also be possible.

In order to evaluate \eq{sexpand}, we need a simple approximation for 
$S$, as a functional of the density. This choice is constrained by a 
number of exact relationships. In a plane-wave representation, 
$S_{\vec q, \vec q\,'}$, one has
the requirements (i) $S_{\vec q,\vec q\,'}(\omega)\rightarrow 
-(4\pi e^2/m\omega^2) n_{\vec q-\vec q\,'}$ at large frequencies (the 
$f$-sum rule), where $n_{\vec k}$ is the Fourier transform of the density; 
(ii) $\int_{-\infty}^{\infty}{du}\; S_{\vec q, \vec q}(iu)
\rightarrow {8\pi^2 N e^2}/{q^2}$ for large $q$, where $N$ is the electron 
number, to reproduce the exactly known self-correlation; 
(iii) $S_{\vec q. \vec q\,'}=S_{-\vec q\,', -\vec q}$ for time reversal invariance;
(iv) a finite $S_{\vec q,\vec q\,'}(\omega)$ for vanishing $q$ or $q'$ at 
all non-zero values of $\omega$, to 
give an exchange-correlation hole with 
the correct volume (charge conservation).

An approximate $S$ inspired by the plasmon-pole model successfully applied
earlier \cite{tractable} takes
$ S_{\vec q, \vec q\,'}=\frac{1}{2}[\tilde S_{\vec q, \vec q\,'}
+  \tilde S_{-\vec q\,', -\vec q} ]$
where
\begin{equation}
\tilde S_{\vec q, \vec q\,'}=
\int\! d^3r\; e^{-i(\vec q - \vec q\,')\cdot \vec r }
\frac{4\pi n(\vec r) e^2/m}{(\omega+\omega_{ q}(\vec r) )
 (-\omega +   \omega_{ q'}(\vec r))}.
\label{tildeSdef}
\end{equation}
We will take $\omega_q(\vec r)$ to be a function of the local density at 
point  $\vec r$ and its gradient. The above $S$ satisfies all the 
constraints provided that $\omega_q \rightarrow q^2/2m$ for large $q$.

To facilitate the numerical evaluation we choose an $\omega_q$ which depends on 
a single length  scale $l$, and switches from its small-$q$ form $\omega_q=1/2ml^2$
to its large-$q$ form (above) when $q\sim 1/l$.
We make an arbitrary choice for the switching function
letting $[1-e^{-(ql)^2}]\omega_q= q^2/2m$. 
The quantity $l$
will be a function of position, and for convenience later
we define $q_0^2 =\gamma/l^2$, where $\gamma=4\pi/9$.
Hence, letting $h(y) = 1-e^{-\gamma y^2}$, we may write
\begin{equation}
\label{eq:dispersion}
\omega_q({\vec r}) = \frac{q^2}{2m} \frac{1}{h(q/q_0({\vec r}))}.
\end{equation}
It would of course be preferable to use a form with a second
length scale, as would be provided by
a linear term $\propto qv_\mathrm{F}$ in the $\omega_q$
dispersion \cite{tractable,rydbergPRL,langrethIJQC},
but this source of error is mitigated by a continuously
variable choice of $q_0$ based on the density.
 We make this choice  so the exchange-correlation
 energy density $\varepsilon^0_{\rm xc}(\vec r)$
 defined by
 \begin{equation}
E_{\rm xc}^0 =\int \! d^3r\;  \varepsilon^0_{\rm xc}(\vec r) n(\vec r)
\label{excrdef}
\end{equation}
produced by this choice of $q_0({\vec r})$ corresponds to that of a full 
calculation.
 
The expression for $E_\text{xc}^0$ corresponding to the approximation
(\ref{ediff2}) for $E_\text{c}^\text{nl}$ is simply
\begin{equation}
E_{\rm xc}^0\approx 
\int_0^\infty\frac{du}{2\pi}\text{tr}\,{(\ln\,\epsilon)}-E_{\rm self}
\approx\int_0^\infty\frac{du}{2\pi}  \rm tr\; S - E_{\rm self}, 
\label{e0a}
\end{equation}
where $E_\text{self}$ subtracts off the internal Coulomb self-energy of
each electron. As was done previously, we expand to lowest order
in $S$ in the second equality.
Substituting for $S$ using   (\ref{tildeSdef}), integrating
over $u=-i\omega$, and using \eq{excrdef} gives 
\begin{equation}
\varepsilon_{\rm xc}^0(\vec r) = \frac{\pi e^2}{m}\int\!\frac{d^3q}{(2\pi)^3}
\left[\frac{1}{\omega_q(\vec r)} -  \frac{2m}{q^2} \right],
\end{equation}
where the second term in the brackets is the self-energy subtraction
written explicitly.
 Upon substitution from (\ref{eq:dispersion}), one finds
\begin{equation}
\varepsilon_{\rm xc}^0(\vec r) =\frac{e^2 q_0(\vec r)}{\pi}\int_0^\infty
\!dy\left[ h(y) -1 \right] = -\frac{3e^2}{4\pi} q_0(\vec r).
\label{excrq0}
\end{equation}
Approximations for $ \varepsilon_{\rm xc}$ are conveniently expressed
as their ratio to the LDA exchange value 
$
\varepsilon_{\rm x}^{\rm LDA} = -{3e^2}k_{\rm F}/{4\pi}, 
$
where $k_{\rm F}^3 = 3\pi^2n$.
Equation~(\ref{excrq0}) then implies that the local value of the
parameter $q_0$ is simply given by the local value of $k_{\rm F}$
modulated by an easily understood energy ratio, that is
\begin{equation}
q_0(\vec r) = \frac{ \varepsilon_{\rm xc}^0(\vec r)}{ 
\varepsilon_{\rm x}^{\rm LDA}%
(\vec r)}
k_{\rm F}(\vec r).
\label{q0value}
\end{equation}

Equation~(\ref{q0value}) is used to determine the $q_0$ value to be 
used in \eq{eq:dispersion}, continuously as a function of position.
For this purpose we use LDA with gradient corrections
\begin{equation}
\varepsilon_{\rm xc}^0\approx \varepsilon_{\rm xc}^\text{LDA}
- \varepsilon_{\rm x}^{\rm LDA}
\left[\frac{Z_\text{ab}}{9}\left(\frac{\nabla n}{2k_\text{F}n}  
  \right)^2\right], 
\end{equation}
where $Z_\text{ab}=-0.8491$. This is the contribution  labeled 
``screened exchange'' in the table in the review of
Ref.~\onlinecite{langrethTrickey}, where the relationship to the original work of 
various authors, who obtained this quantity from first principles, 
is discussed.  The remaining contribution $Z_\text{c}$, labeled ``fluctuation''
in Ref.~\onlinecite{langrethTrickey}, 
is part of $E^\text{nl}_\text{c}$, and hence inappropriate to include as part 
of $E_\text{xc}^0$. It comes from a 
term in perturbation theory that gives a long range vdW-like 
interaction if taken at long range instead of in an expansion in gradients
\cite{fluctuation}.

\begin{figure}[ht]
\centerline{\includegraphics[width=0.41\textwidth]{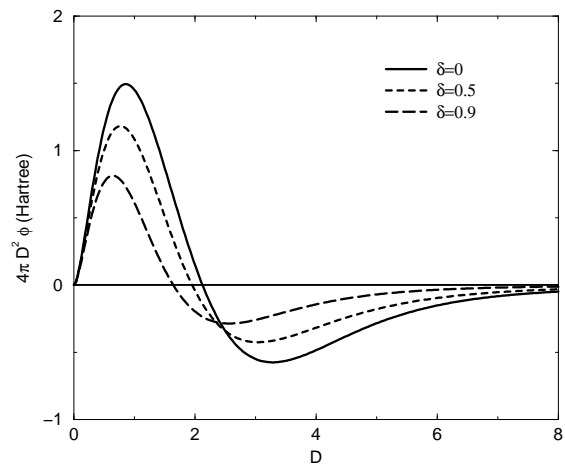}}
\caption{\label{fig:phiplot}
The kernel $\phi$ in \eq{twopoint} 
as a function of the dimensionless $D$ parameter for several values of 
the asymmetry parameter $\delta$, as defined in the text.
}
\end{figure}

Writing \eq{sexpand} in a plane-wave representation gives
\begin{equation}
E_{\rm xc}^\text{nl}= \int _0^\infty \frac{du}{4\pi}\sum_{\vec q, \vec q\,'}
\left[ 1 - {(\hat q \cdot  \hat q')^2} \right] S_{\vec q, \vec q\,'}
S_{\vec q\,', \vec q}.
\label{sexpandf}
\end{equation}
This may be straightforwardly, but tediously, 
expressed in the form of \eq{twopoint}, where the kernel $\phi$ is given by
\begin{eqnarray}
\phi(\vec r, {\vec r}\,')&=&\frac{2me^4}{\pi^2}
\int_0^\infty\!a^2\,da\int_0^\infty\!b^2\,db\, W(a,b) \nonumber \\
&\times&T\left(\nu(a), \nu(b),
\nu'(a),\nu'(b)\right),
\label{phi3}
\end{eqnarray}
where 
\begin{eqnarray}
T(w,x,y,z)&=&
\frac{1}{2}\left[\frac{1}{w+x}+\frac{1}{y+z}\right]   \\
&&\times\left[\frac{1}{(w+y)(x+z)}
+ \frac{1}{(w+z)(y+x)}\right] \nonumber
\label{Teval}
\end{eqnarray}
and
\begin{eqnarray}
&{W(a,b)=
{2}[(3-a^2)b\cos b \sin a 
+(3-b^2)a \cos a \sin b} \nonumber \\
&+ (a^2+b^2-3)\sin a \sin b -3ab \cos a \cos b]/a^3b^3.
\label{Weval}
\end{eqnarray}
The quantities $\nu$ and $\nu'$ are given by 
$\nu(y)={y^2}/{2h(y/d)}$ and 
$\nu'(y)={y^2}/{2h(y/d')}$,
with
$d= |\vec r - \vec r\,' |q_0(\vec r) $ and  $d'= 
 |\vec r - \vec r\,' |  q_0(\vec r\,')$,
where $q_0$  is given by
Eq.~(\ref{q0value}).  The kernel $\phi$ thus depends on $\vec r$ and
$\vec r\,'$  only through $d$ and $d'$, so that $\phi$ can be tabulated
in advance in terms of these two variables, or better yet in terms
of the sum and difference variables $D$ and $\delta$ defined by 
$d=D(1+\delta)$ and $d'=D(1-\delta)$.  Then $0\le D < \infty$
and $0\le |\delta| < 1$.  For large $d$ and $d'$, the asymptotic form is
\begin{equation}
\phi
\rightarrow
-\frac{C}{d^2d'^2\left(d^2+d'^2\right)},
\label{phiasymp}
\end{equation}
where 
$C=12(4\pi/9)^3me^4$.  In Fig.~\ref{fig:phiplot} we show a plot of 
$4\pi D^2 \phi$
vs.~$D$ for several values of $\delta$. The integral of the
$\delta=0$ curve vanishes as it must.

The numerical  work uses \eq{ec0} for the correlation functional
coupled with  the Zhang-Yang revPBE \cite{revPBE} exchange functional.
This choice is motivated by the work of Wu {\it et al.}\ \cite{scoles},
which 
pointed out that a more standard GGA 
predicts substantial binding in rare gas dimers from exchange alone, a 
feature absent for exact Hartree-Fock exchange.  We 
found \cite{langrethIJQC} that revPBE exchange 
does not have this property, 
so by using it we insure that vdW binding, a correlation effect, actually 
comes from the correlation term in our approximation scheme \cite{method}. 
 
In Table \ref{c6} our 
 calculated values for the 
coefficient $C_6$ in the asymptotic interaction $-C_6/R^6$ 
between the elements of  several dimers are compared with 
those from  previous related calculations as well as  
reference values, which also have some uncertainty. 
The new values track those from Ref.~\onlinecite{ylvaPRL} which 
used the same approximation for the electrodynamics, and are only slightly 
inferior to the values obtained in Ref.~\onlinecite{unified}, where 
self-consistent electrodynamics 
was used. 

\begin{figure}[ht]
\centerline{\includegraphics[width=0.41\textwidth]{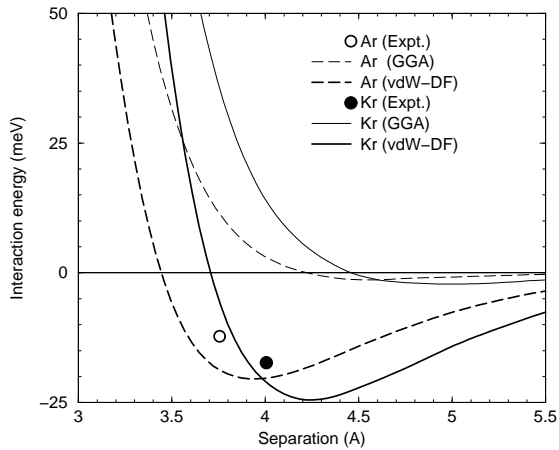}}
\caption{\label{fig:Ar-Kr}
Calculated interaction energy between two Ar atoms (dashed curves)
and between two Kr atoms (solid curves). The experimental equilibrium
data \cite{Ar-KrRef} are shown for comparison with the full vdW-DF,
and the GGA predictions in the revPBE flavor are also shown.
}
\end{figure}

\begin{figure}[ht]
\centerline{\includegraphics[width=0.41\textwidth]{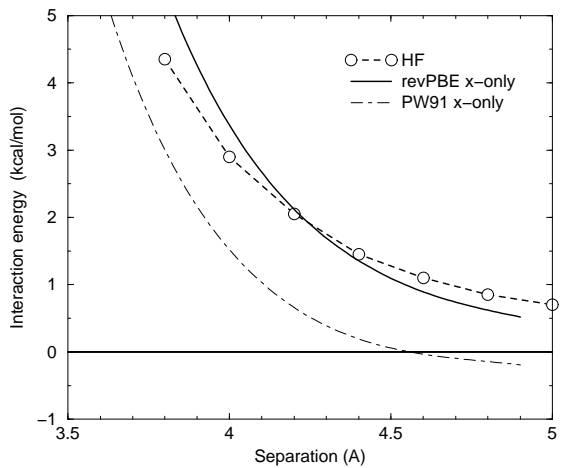}}
\caption{\label{fig:benzene_x}
Interaction energy between two benzene molecules
 in the atop-parallel configuration as predicted by the exchange-only
 part of two GGA functionals.  The full Hartree-Fock results
 \cite{CCSD(T)MP2} are shown for comparison.
 }
 \end{figure}

\begin{figure}[ht]
\centerline{\includegraphics[width=0.41\textwidth]{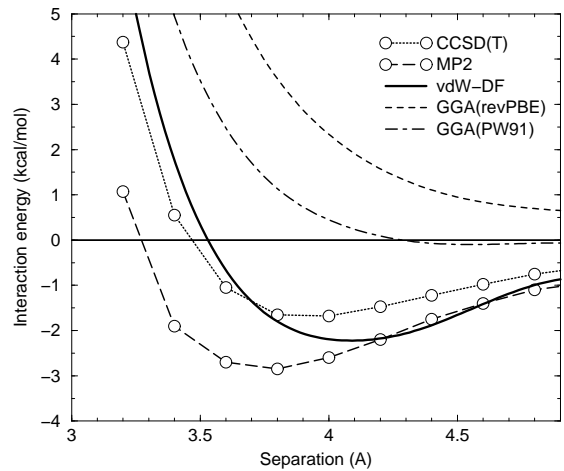}}
\caption{\label{fig:benzene}
Full interaction energy between two benzene molecules
 in the atop-parallel configuration using the vdW-DF functional. 
For comparison, we show recent results \cite{CCSD(T)MP2}
using 
coupled-cluster [CCSD(T)] and 
perturbation-theoretic  
(MP2) methods, as well as the prediction of two flavors of GGA.
Our vdW-DF would have given an equilibrium separation
closer to those from the 
wavefunction calculations,
if exact HF exchange (see Fig.~\ref{fig:benzene_x}) had been
used instead of revPBE exchange.
}
\end{figure}

Figure ~\ref{fig:Ar-Kr} shows the calculated binding-energy curves 
as functions of separation   
for Ar  
and 
Kr dimers. The comparison with 
experimental values for the binding 
energy and distance~\cite{Ar-KrRef} illustrate 
the promise of vdW-DF for such systems.
The 
binding-energy curves for a benzene dimer in the 
atop-parallel configuration  
in Fig.~\ref{fig:benzene_x} 
illustrate 
the importance of choosing the right GGA flavor, \textit{i.e}. revPBE~\cite{revPBE}, 
to avoid erroneous attraction in exchange-only accounts, as discussed in 
Refs.~\onlinecite{langrethIJQC} and \onlinecite{scoles}. 
Figure~\ref{fig:benzene} illustrates
the relative agreement between 
modern 
wavefunction based 
calculations~\cite{CCSD(T)MP2}  
and our vdW-DF method,
and the importance of
not using the GGA alone.

The above moderate successes of the vdW-DF proposed here suggest
that its use, along with future improvements, may be a way to proceed
for calculating properties of vdW bound molecules that are too large
for wavefunction based methods to be useful.

Financial support from the Swedish Foundation for Strategic Research 
via Materials Consortia \#9 and ATOMICS and the Swedish Scientific 
Council 
is gratefully acknowledged. Work by M.D.\ and D.C.L.\ 
was supported in part by NSF Grant DMR 00--93070.

\begin{table}[ht]
\caption{\label{c6}
$C_6$ values  for  dimers (Rydberg atomic units). Present: from \eq{phiasymp}.
Similar: calculations from Ref.~\onlinecite{ylvaPRL} using the 
same electrodynamics
approximation. Unified: calculations from Ref.~\onlinecite{unified} using 
self-consistent electrodynamics. Reference: sources cited in
Ref.~\onlinecite{unified}. 
}
\begin{tabular}{lcccc}
\hline\hline
Dimer		& Present	& Similar	& Unified 	& Reference \\
\hline
He		&4.8		&4		&2.58		&2.92	\\
Ne		&14.6		&12		&15.0		&13.8	\\
Ar		&124		&126		&143		&134	\\
Kr		&238		&245		&291		&266	\\
Xe		&516		&520		&663		&597	\\
Mg		&1598		&1513		&1230		&1240	\\
\hline
\end{tabular}
\end{table}


\begin{thebibliography}{10}
\bibitem{fluctuation}
Alternatively, an expansion of \eq{sexpand} in gradients gives the precise
  structure of $Z_\text{c}$, but with the triangular vertex parts [see
  Fig.~3(c) in Ref.~\onlinecite{langrethTrickey}] evaluated in local-density
  approximations. The gross failure of the old-fashioned gradient
  approximation, due to the unwitting attempt to expand the van der Waals
  interaction in terms of semilocal gradients, is obviated here by treating the
  van der Waals term in a more appropriate approximation.
\bibitem{method}
The calculations were done with the {\sc dacapo} plane-wave pseudopotential
  code, using the full revPBE exchange-correlation potential.
  $E_\text{c}^\text{nl}$ was added and the beyond LDA part of the revPBE
  correlation energy subtracted as post GGA perturbations. The core densities
  were restored by hand and found to make a negligible difference in
  $E_\text{c}^\text{nl}$.
\end{thebibliography}

\end{document}